\documentclass[10pt,letterpaper,twocolumn]{article} 

\usepackage{ol2}
\usepackage[draft]{hyperref}
\usepackage{amsmath}

\begin{document}

\twocolumn[ 

\title{Delocalization of light in photonic lattices with unbounded potentials}


\author{Stefano Longhi}
\address{Dipartimento di Fisica, Politecnico di Milano and Istituto di Fotonica e Nanotecnologie del Consiglio Nazionale delle Ricerche, Piazza L. da Vinci 32, I-20133 Milano, Italy (stefano.longhi@polimi.it)}
\address{IFISC (UIB-CSIC), Instituto de Fisica Interdisciplinar y Sistemas Complejos, E-07122 Palma de Mallorca, Spain}

\begin{abstract*}
In classical mechanics, a particle cannot escape from an unbounded potential well. Naively, one would expect a similar result to hold in wave mechanics, since high barriers make tunneling difficult. However, this is not always the case and it is known that wave delocalization can arise in certain models with incommensurate unbounded potentials sustaining critical states, i.e. states neither fully extended nor fully localized. Here we introduce a different and broader class of unbounded potentials, which are not quasi-periodic and do not require any specially-tailored shape, where wave delocalization is observed. The results are illustrated by considering light dynamics in synthetic photonic lattices, which should provide a feasible platform for the experimental observation of wave delocalization in unbounded potentials.
\end{abstract*}

 ] 

{\em Introduction.} 
Transport inhibition via Anderson localization \cite{r1} is ubiquitous in wavy systems with uncorrelated static disorder or aperiodic order. Wave function localization arises from the delicate destructive interference
among multiply scattered waves, which localizes all wave functions and prevents transport in the system. This kind of localization has been observed in a wide variety of physical systems, including photonic systems (see e.g. \cite{r2,r3,r4,r5} and references therein ). 
In classical mechanics, a simple way to localize in space a particle is to create a potential well of infinite depth, like an harmonic oscillator, so as the particle cannot escape from the well regardless of its energy. It is quite intuitive to think that the same localization mechanism, which is very distinct than Anderson localization, should persist in wave mechanics, given that  high barriers make tunneling difficult and prevent at the end wave delocalization. This is of course the case when the potential is monotonously increasing at infinity; however more subtle effects can arise when the potential is unbounded but oscillates between high and low values, in such a way that the wave function alternates between evanescent and propagative regions. Dynamical wave delocalization is commonly related to the existence of extended wave functions in the system, i.e. to the absolutely continuous spectrum of the Hamiltonian. According to the Simon-Spencer theorem \cite{r6}, the absolutely continuous spectrum for Hamiltonians with  unbounded potentials is empty, and thus one might conclude that transport is impossible when dealing with unbounded potentials, even when they are oscillating. For example, localization is known to arise in unbounded monotone and quasi-periodic potentials on a lattice \cite{r7},  such as in the Maryland model introduced in the context of quantum chaos and dynamical localization \cite{r8,r9,r10,r11,r12,r13}. However, recent studies \cite{r14,r15,r16,r17} unravelled that in other unbounded potentials with aperiodic order not all the wave functions are fully localized, and a large fraction of them are critical states, i.e. they are neither fully localized nor fully extended. In such special unbounded potentials, dynamical delocalization is possible and mediated by the critical (rather than the extended) states, generally resulting in diffusive or anomalous diffusive (rather than ballistic) transport \cite{r17b}.\\
In this Letter it is shown that dynamical wave delocalization can be observed rather universally in a wide class of unbounded potentials with an almost pure point energy spectrum, and that delocalization does not necessarily require any kind of aperiodic order in the system nor any specific shape of the potential. Wave transport arises here from the existence of bands of weakly-localized states (rather than critical states), emanating from isolated fully extended states. 
 We illustrate such a result by considering delocalization of light in photonic lattices, and suggest  discrete-time synthetic mesh lattices as a suitable platform for the observation of wave delocalization in unbounded oscillating potentials.\\
\begin{figure}[ht]
    \includegraphics[width=0.45\textwidth]{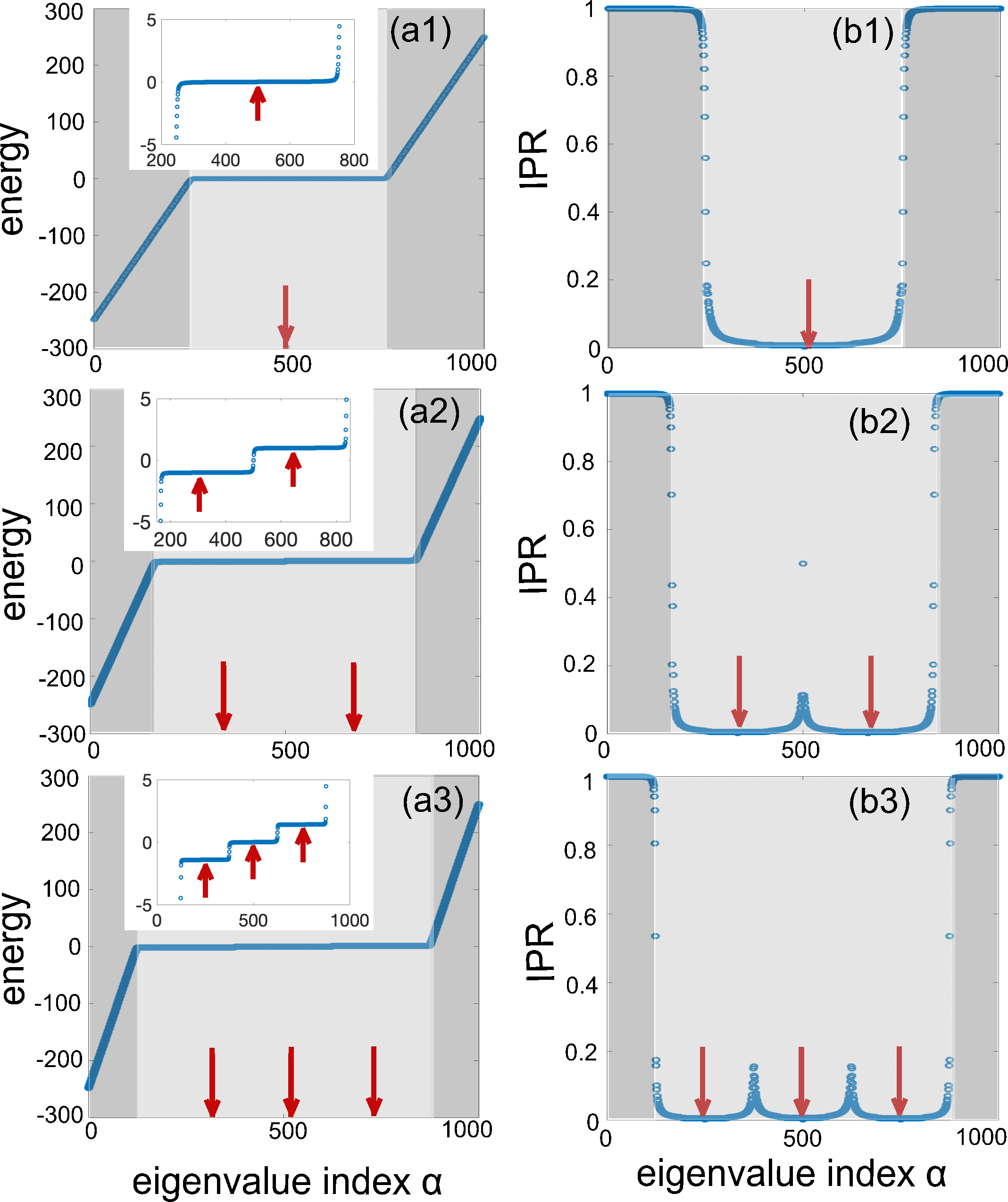}
    \caption{Behavior of energy  {\color{black} [panels (a1), (a2) and (3)]} and IPR  {\color{black} [panels (b1), (b2) and (b3)]} of the $L$ eigenfunctions of the Hamiltonaina $\mathcal{H}$ in a finite lattice of size $L=1000$ for a linear potential $f_l=Fl$ ($F=0.5 \kappa$)  and for a few increasing values of $M$. {\color{black} (a1) and (b1)}: $M=2$; {\color{black} (a2) and (b2)}: $M=3$;  {\color{black} (a3) and (b3)}: $M=4$. The energy is in units of $\kappa$. The dark shaded areas correspond to strongly localized states, the light dashed areas correspond to weakly localized states. The vertical bold arrows mark the isolated extended states. The insets {\color{black} in panels (a1), (a2) and (a3)} show an enlargement of the spectrum for the low-energy and weakly-localized wave functions, which form $(M-1)$ energy plateaus.}
\end{figure}
 \begin{figure}[ht]
    \includegraphics[width=0.45\textwidth]{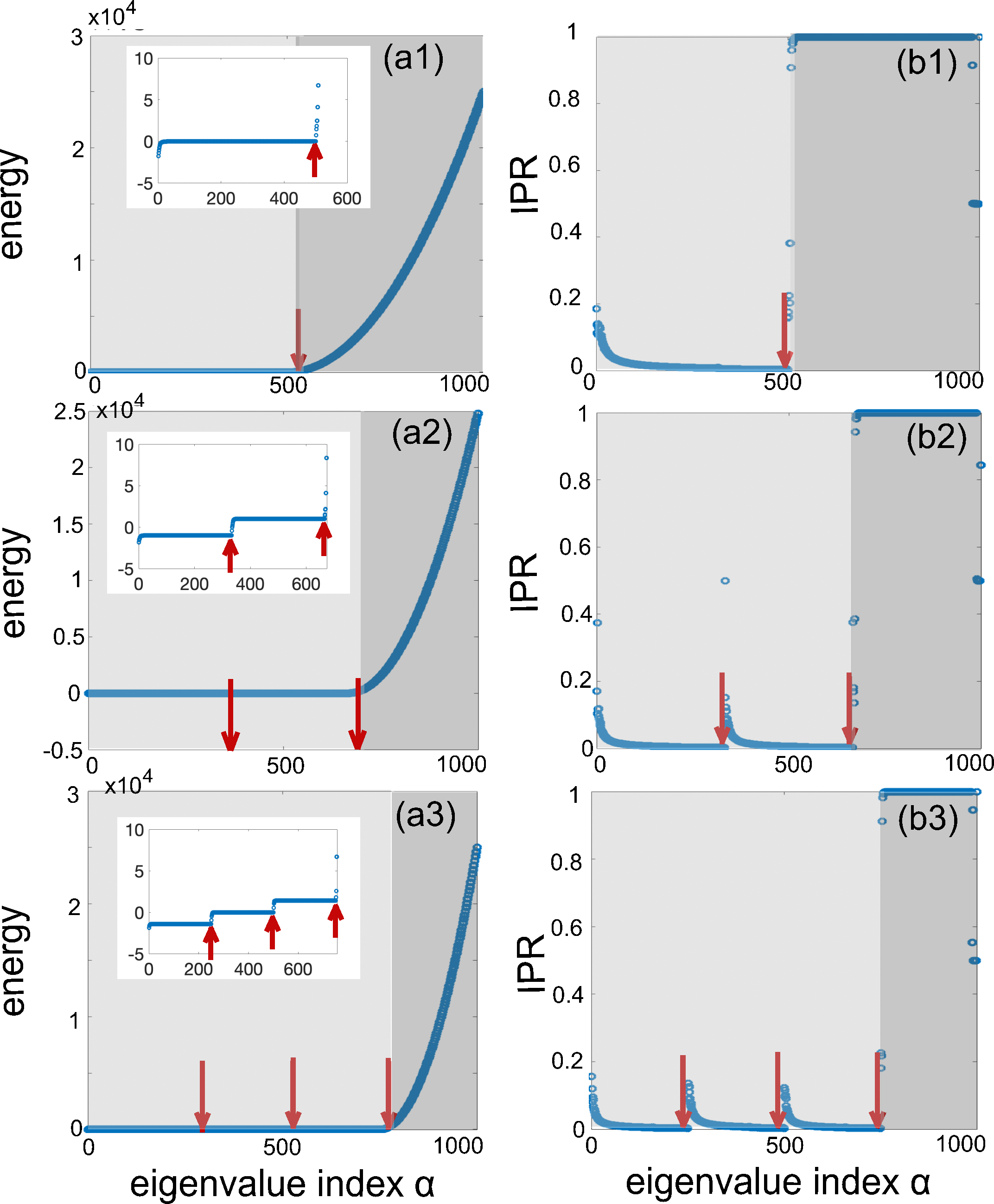}
   \caption{Same as Fig.1, but for a parabolic potential $f_l=Fl^2$ with $F=0.1 \kappa$.}
 \end{figure}

{\it Unbounded potential on a lattice: model and spectral analysis.}  We consider discretized light dynamics on a one-dimensional photonic lattice, such as arrays of evanescently-coupled optical waveguides or synthetic lattices in frequency or time domains (see e.g. \cite{r18,r19,r20}). Assuming nearest-neighbor coupling, light dynamics is governed by the set of coupled equations \cite{r18,r19,r20,r21}
\begin{equation}
 i \frac{d c_n}{dt}= \kappa(c_{n+1}+c_{n-1})+V_n c_n, 
 \end{equation}
 where $\kappa$ is the coupling constant and the potential $V_n$ describes propagation constant detuning (or resonance detuning) from a reference value. The eigenfunctions of the lattice, $c_n= \psi_n \exp(-iEt)$, satisfy the eigenvalue equation
\begin{equation}
E \psi_n= \kappa(\psi_{n+1}+\psi_{n-1})+ V_n \psi_n \equiv \mathcal{H} \psi_n
\end{equation}
with Hamiltonian $\mathcal{H}$. Indicating by $f_n$ a rather arbitrary function of the discrete variable $n$, satisfying the minimal requirements that $|f_n|$ is asymptotically a monotonously-increasing function of $n$ with $\lim_{n \rightarrow \pm \infty} |f_n|=\infty$ and $f_0=0$, let us assume as a potential $V_n$ the oscillating and unbounded function defined by
\begin{equation}
V_n= \left\{
\begin{array} {l l}
f_n  & n=0, \pm M, \pm 2M, \pm 3 M, ... \\
0 & \;  {\rm otherwise} 
\end{array}
\right.
\end{equation}
where $M$ is an assigned integer number. Clearly, for $M=1$ one simply has $V_n=f_n$, and all the eigenfunctions $\psi_n$ are fully localized (normalizable) since $|V_l |$ is a monotonously increasing function as $n \rightarrow \pm \infty$. Remarakly, for $M \geq 2$, i.e. when the potential oscillates between large values and zero, this is not the case, and extended wave functions do exist. Specifically,  it can be readily shown that the set of $(M-1)$ wave functions
\begin{equation}
\psi_n^{(\sigma)}= \sin (\pi \sigma n/M )
\end{equation}
with energies 
\begin{equation}
E_{\sigma}=2 \kappa \cos (\pi \sigma /M)
\end{equation}
($\sigma=1,2,..,M-1$), corresponding to fully delocalized states, are non-normalizable eigenfunctions of Eq.(2). We note that such a result is not at odd with the Simon-Spencer theorem \cite{r6}, since the extended states correspond to isolated points in energy spectrum with zero spectral measure. However, the appearance of extended states would suggest that, besides strongly localized and isolated extended wave functions, there should arise bands of weakly-localized wave functions that could effectively enable transport in the lattice. This  conjecture is confirmed by numerical diagonalization of the matrix Hamiltonian $\mathcal{H}$, assuming a finite lattice of large size $L$ with periodic boundary conditions. Rather generally, besides the $M$ isolated extended states given by Eq.(4), it turns out that a number of $ \sim L/M$ wave functions with 'high' energies (larger in modulus than $\sim 2 \kappa$) are strongly localized, whereas a number of $ \sim L(1-1/M)$ wave functions with 'low' energies (in modulus smaller than $ \sim 2 \kappa$) display a weak localization, approaching a fully extended state as their energy gets close to one of the values $E_{\sigma}$ defined by Eq.(5).
 Intuitively, such a result can be explained as follows. For a given large integer $n_0$, which is an integer multiple than $M$ such that $|V_{n_0}| \gg 2 \kappa$, excitation can be trapped at the impurity site $n=n_0$, which is not resonant with neighboring sites in the lattice; correspondingly, the wave function $\psi_n$ is tightly confined near $n=n_0$ and the eigenenergy is approximately equal to the on-site potential energy $\sim V_{n_0}$. On the other hand, excitation at any site $n_0$, with $n_0$ prime with respect to $M$ and corresponding to a zero on-site potential $(V_{n_0}=0$), is separated by alternating high barriers from adjacent resonant sites, and thus tunneling could arise all along the lattice, which would explain the possibility of transport. To illustrate such results, we will focus our attention by considering two simple potentials: the linear potential, $f_l=Fl$,  and the parabolic potential, $f_l=Fl^2$, {\color{black} however the results hold for rather arbitrary potential profiles, 
 not necessarily with odd or  even symmetry for spatial inversion and with an asymptotic growth different than polynomial (see the Supplemental document)}. For $M=1$, the energy spectrum of $\mathcal{H}$ is pure point, all the wave functions are fully localized and transport in the lattice is prevented. In particular, it is well known that for the linear potential $V_l=Fl$ the energy spectrum forms a Wannier-Stark ladder and the dynamics is periodic in time, leading to the famous Bloch oscillations which have been observed in several photonic settings (see e.g. \cite{r21,r22,r23,r24}). For the parabolic potential $V_l=F l^2$, the energy spectrum is pure point but not equally spaced and the dynamics displays a transition from dipolar to Bloch oscillations \cite{r25,r26}.\\		 
The localization properties of a wave function $\psi_n^{(\alpha)}$, normalized as $ \sum_{n=1}^{L} | \psi_n^{(\alpha)}|^2=1$, are characterized by the inverse
participation ratio (IPR), defined by
\begin{equation}
 IPR_{\alpha}=  \sum_{n=1}^{L} | \psi_n^{(\alpha)} |^4
 \end{equation}
 (with $IPR \leq 1$),
  and by the fractal dimension $\beta_{\alpha}$, defined by
  \begin{equation}
  \beta_{\alpha}=\lim_{L \rightarrow \infty} \frac{\log (IPR_{\alpha})}{\log(1/L)}.
  \end{equation}
  {\color{black} The $IPR$ and fractal dimension are generally used to quantify the spatial extent or localization of the wavefunctions in a disordered system, and provide a useful tool for characterizing the transition between localized, critical  and extended states. For a localized eigenstate, the $IPR$ takes a finite value independent of the system size $L$ in the large $L$ limit, which corresponds to a vanishing fractal dimension $\beta_{\alpha}=0$. A weakly (strongly) localized eigenstate is characterized by a small (large) value of the $IPR$, and in both cases $\beta_{\alpha}=0$. Conversely, for extended and critical states the IPR vanishes as $L \rightarrow \infty$, but with  a different decay law versus $L$ yielding distinct fractal dimensions
   $\beta_{\alpha}=1$ for an uniformly extended state, and $0<\beta_{\alpha}<1$ for a critical state \cite{r16,r17,r27}.}  
  Figures 1 and 2 show, as examples, the numerically-computed energy spectrum and corresponding values of $IPR$ of all eigenfunctions in a finite lattice comprising $L=1000$ sites for the linear (Fig.1) and parabolic (Fig.2) potentials and for a few increasing values of $M$. {\color{black} The behavior of the $IPR$ clearly shows the coexistence of strongly localized eigenstates (the dark shaded regions in the plots with large values of $IPR$) and weakly localized eigenstates 
  (the light shaded regions in the plots corresponding to low values of $IPR$)}. Note that the fraction of weakly-localized states versus strongly-localized states is given by $ \sim (M-1)$.
  {\color{black} The isolated extended states, at energies $E_{\sigma}=2 \kappa \cos (\pi \sigma /M)$, are indicated by the vertical bold arrows in Figs.1 and 2, and are embedded in the regions of 
  weakly-localized states, at the center or edges of the $(M-1)$ energy plateaus of the weakly-localized states, as shown in the insets 
  of panels (a1), (a2) and (a3) in Figs.1 and 2}.  A detailed fractal analysis, which is presented in the Supplemental document, indicates that the wave functions at energies $E$ close to one of the $(M-1)$ values $E_\sigma$ are weakly localized rather than critical states, and they become more and more delocalized as $E$ approaches $E_{\sigma}$. This means that, contrary to unbounded potentials with quasiperiodic order considered in recent works \cite{r14,r15,r16,r17}, in our class of unbounded potentials the energy spectrum is almost pure point and there are not critical states nor mobility edges, separating localized and critical states \cite{LonghiWS}.\\ 
\\
{\it Unbounded potential on a lattice: wave delocalization and photonic implementation.}
The appearance of a band of weakly localized states, with a diverging localization near the isolated energies $E_{\sigma}$ corresponding to the fully extended states, can enable wave delocalization and transport in the lattice, in spite of the high potential barriers of the unbounded potential and the almost pure point spectrum of the Hamiltonian \cite{r27b}. Since the absolutely continuous spectrum is empty, transport is rather generally diffusive or anomalous diffusive rather than ballistic \cite{r27b,r28}. In experiments, it can be characterized  by measuring, as a function of time, some dynamical variables such as the second-order moment of the spreading wave packet or the dynamic IPR for an initial single-site excitation of the lattice (see for instance \cite{r29,r30,r30b}). Light delocalization could be observed  using different optical platforms, such as waveguide arrays with suitably-tailored waveguide size and spacing \cite{r21}, synthetic lattices in frequency space \cite{r20,r24} or pulse dynamics in synthetic mesh optical lattices \cite{r23,r30,r31,r32,r33,r34}. Here we illustrate light delocalization in the latter photonic platform. 
The system  consists of two fiber loops of slightly different lengths that are connected by a fiber coupler with a coupling angle $\beta$ {\color{black}(see Supplemental document)}. A phase modulator is placed in one of the two loops, which provides a desired control of the phase  of the traveling pulses. 
Light dynamics is described by the discrete-time coupled-mode equations \cite{r23,r30,r31,r32,r33,r34}
 \begin{eqnarray}
 u^{(m+1)}_n & = & \left(   \cos \beta u^{(m)}_{n+1}+i \sin \beta v^{(m)}_{n+1}  \right)  \exp (-2i\phi_n) \; \\
 v^{(m+1)}_n & = & \left(   \cos \beta v^{(m)}_{n-1}+i \sin \beta u^{(m)}_{n-1}  \right)
 \end{eqnarray}
 \begin{figure}[ht]
  \centering
    \includegraphics[width=0.49\textwidth]{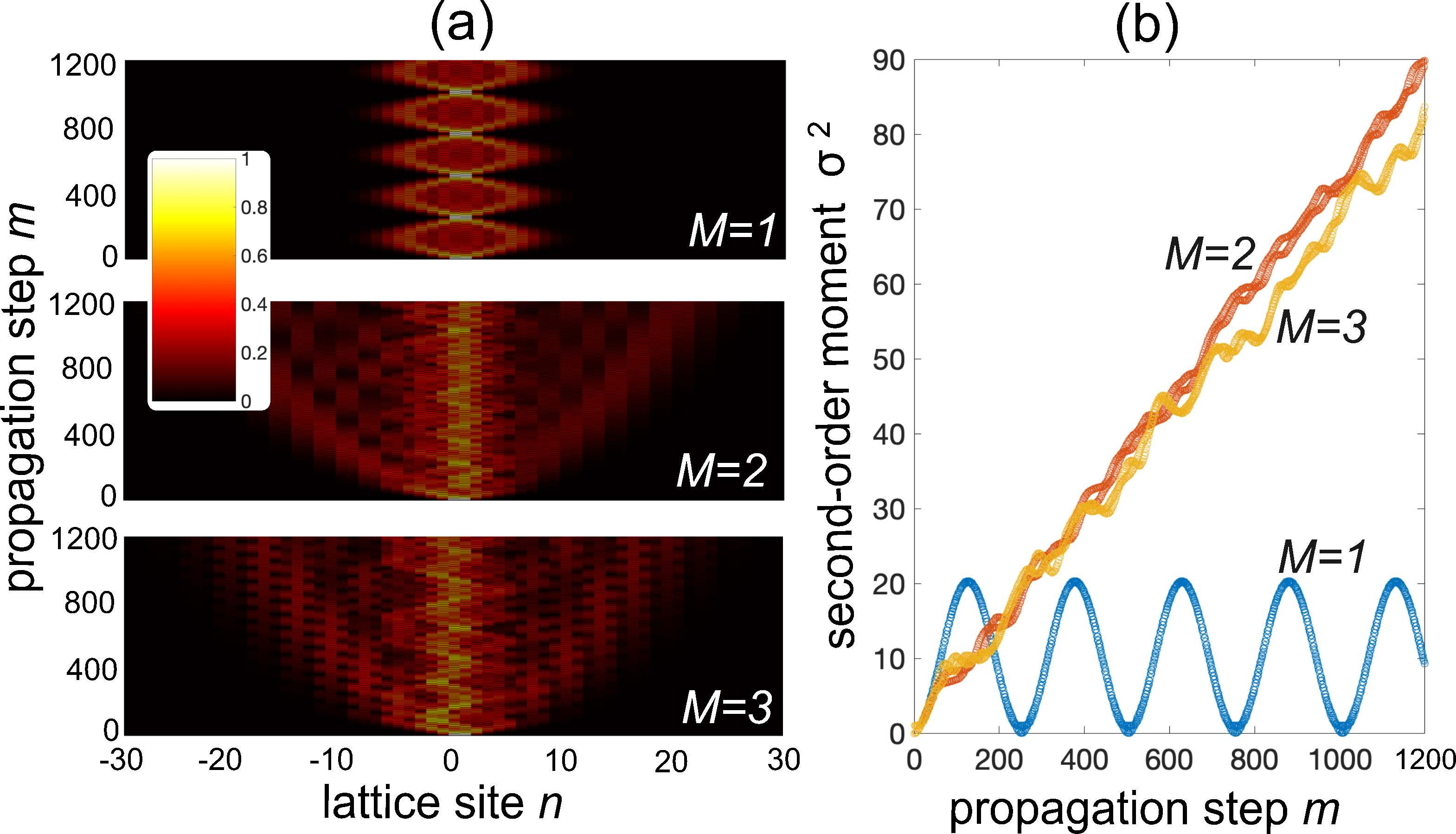}
   \caption{(a) Light dynamics in a synthetic mesh lattice for $\beta= 0.95 \times \pi /2$, a linear potential $f_n=Fn$ ($F=0.025$) and for $M=1,2,3$. (b) Corresponding behavior of the second-order moment $\sigma^2(m)$. Initial excitation of the lattice is 
   $u_n^{(0)}= \delta_{n,0}$ and $v_n^{(0)}=0$.}
\end{figure}
 \begin{figure}[ht]
  \centering
    \includegraphics[width=0.49\textwidth]{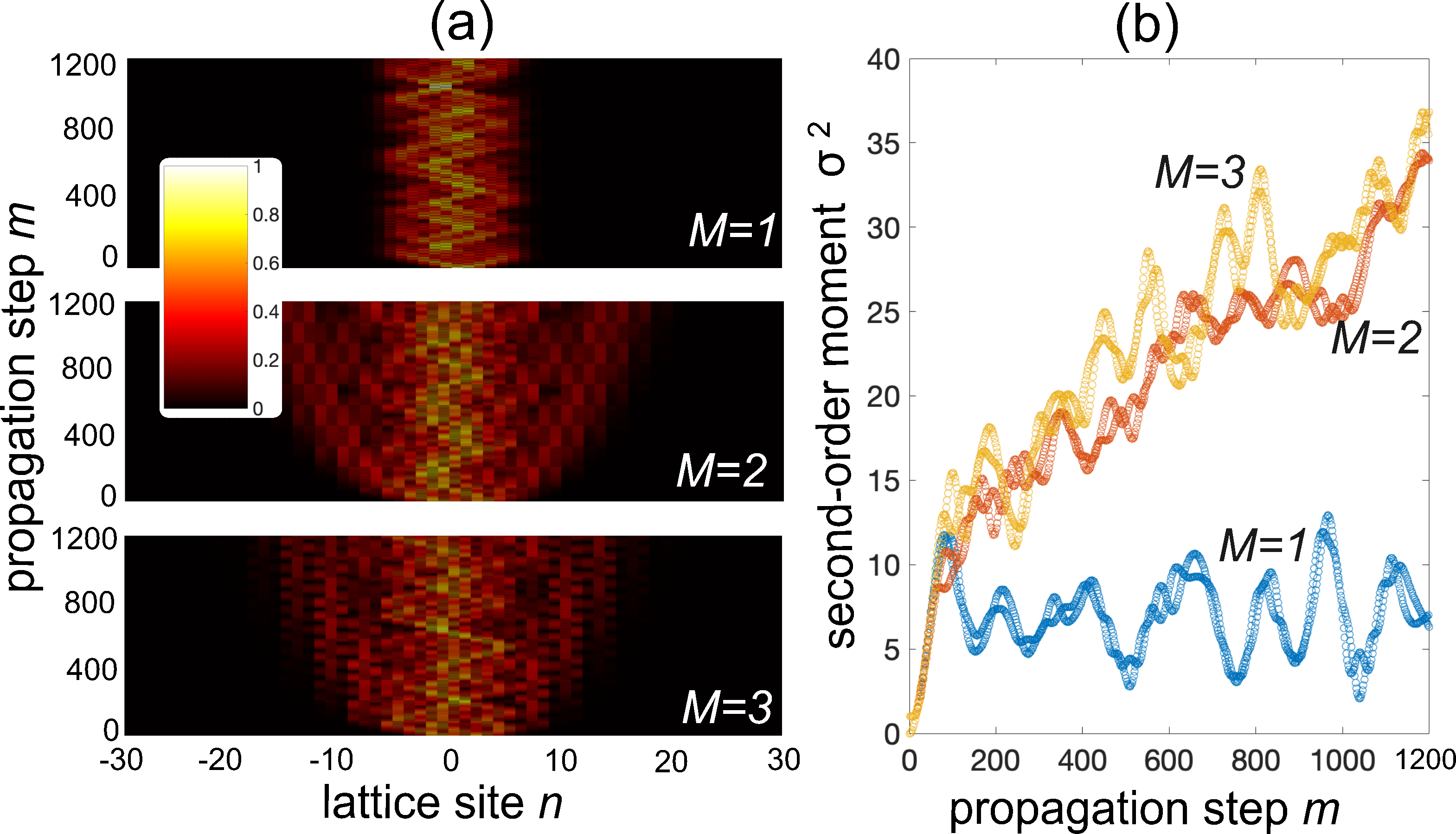}
   \caption{Same as Fig.3, but for a parabolic potential $f_n=Fn^2$ ($F=0.05$).}
\end{figure}
where $u_n^{(m)}$ and $v_n^{(m)}$ are the pulse amplitudes at discrete time step $m$ and lattice site $n$ in the two fiber loops, and $2 \phi_n$ is the phase term. Assuming a coupling angle $ \beta$ close to $\pi /2$ and for a weak phase modulation, the light dynamics can be effectively described by the continuous-time model Eq.(1), with the discrete time $m$ replaced by a continuous time variable $t$ and with $\kappa= (1/2) \cos\beta$, $V_n= \phi_n$; technical details are given in the Supplemental document (see also \cite{r34,r34b}). Wave spreading in the lattice is monitored by the time evolution of the second-order moment $\sigma^2(m)=\sum_n n^2 (|u_{n}^{(m)}|^2+|v_{n}^{(m)}|^2)$ when the lattice is initially excited in the single site $n=0$. Assuming asymptotically $\sigma^2(m) \sim m^{2 \delta}$, the exponent $\delta$ characterizes the kind of transport, with $\delta=1$ for ballistic transport, $\delta=1/2$ for diffusive transport and $\delta=0$ for dynamical localization (see e.g. \cite{r30}). The impressed phase term $\phi_n=V_n$ is tailored according to Eq.(3), with $f_n$ either a linear or a parabolic function of $n$. Figures 3 and 4 show typical light dynamics in the synthetic lattice for a linear (Fig.3) and parabolic (Fig.4) potentials and for a few increasing values of $M$. Clearly, for $M=1$ one has localization, with the characteristic Bloch oscillation dynamics in the linear potential case with a Bloch oscillation period given by $2 \pi /F$ (Fig.3). Such a revival regime has been observed in previous experiments \cite{r23,r35}, and rigorous analysis was presented in \cite{r36}. In the parabolic potential case, we still observe  dynamical localization, however the dynamics is not periodic owing to the non-equally spacing of the energies for the discrete parabolic potential. Conversely, delocalization is observed for $M=2$ and 3, both for linear and parabolic potentials, with a spreading exponent $\delta$ close to 1/2, i.e. with a linear increase of $\sigma^2(m)$ with $m$ [see Figs.3(b and 4(b)], indicating 
a nearly-diffusive transport. {\color{black} An additional example of wave delocalization in a potential well with exponential (rather than polynomial) growth, lacking of any symmetry for spatial inversion, is presented in 
the Supplemental document.}\\
\\ 
 {\it Conclusion.} We predicted dynamical delocalization of waves in a broad class of unbounded potentials on a lattice, displaying an almost pure point spectrum and without mobility edges. Unlike delocalization in unbounded incommensurate potentials predicted in recent works, here transport is mediated by bands of weakly localized (rather than critical) states, and special tailoring of the potential is not required. Our results shed new physical insights onto the localization properties of unbounded potentials, and indicate that light dynamics in synthetic mesh lattices could provide an experimentally accessible platform for the observation of localization-delocalization transitions.\\
\\
\noindent
{\bf Disclosures}. The author declares no conflicts of interest.\\
{\bf Data availability}. No data were generated or analyzed in the presented research.\\
{\bf Supplemental document}. See Supplement 1 for supporting content.

\newpage


 {\bf References with full titles}\\
 \\
 \noindent
 1. P.W. Anderson,  Absence of Diffusion in Certain Random Lattices  Phys. Rev. {\bf 109},1492 (1958).\\
2. D.S. Wiersma, P. Bartolini, A. Lagendijk, and R. Righini, Localization of light in a disordered medium, Nature {\bf 390}, 671 (1997).\\
3. T. Schwartz, G. Bartal, S. Fishman, and M. Segev, Transport and Anderson localization in disordered two-dimensional photonic lattices, Nature {\bf 446}, 52 (2007).\\
4. Y. Lahini, R. Pugatch, F. Pozzi, M. Sorel, R. Morandotti, N. Davidson, and Y. Silberberg, Observation of a Localization Transition in Quasiperiodic Photonic Lattices, Phys. Rev. Lett.
{\bf 103}, 013901 (2009).\\
5.  M. Segev, Y. Silberberg, and D.N. Christodoulides, Anderson localization of light, Nature Photon. {\bf 7}, 197 (2013).\\
6. B. Simon and T. Spencer, Trace Class Perturbations and
the Absence of Absolutely Continuous Spectra, Commun.
Math. Phys. {\bf 125}, 113 (1989).\\
7. I. Kachkovskiy, Localization for quasiperiodic operators with unbounded monotone potentials, J. Functional Analysis {\bf 277}, 3467 (2019).\\
8. D.B. Grempel, S. Fishman, and B. E. Prange, Localization in an Incommensurate Potential: An Exactly Solvable Model, Phys.
Rev. Lett. {\bf 49}, 833 (1982).\\
9. S. Fishman, D.B. Grempel, and B. E. Prange, Chaos,
Quantum Recurrences, and Anderson Localization, Phys. Rev.
Lett. {\bf 49}, 509 (1982).\\
10. B. Simon, Almost Periodic Schr\"odinger Operators IV. The
Maryland Model, Ann. Phys. {\bf 159}, 157 (1985).\\
11. B. Fischer, B. Vodonos, S. Atkins, and A. Bekker, Experimental
demonstration of localization in the frequency domain of
mode-locked lasers with dispersion, Opt. Lett. {\bf 27}, 1061 (2002).\\
12. S. Ganeshan, K. Kechedzhi, and S. Das Sarma, Critical
integer quantum Hall topology and the integrable Maryland
model as a topological quantum critical point, Phys. Rev. B {\bf 90},
041405(R) (2014).\\
13. S. Longhi, Maryland model in optical waveguide lattices, Opt. Lett. {\bf 46}, 637 (2021).\\
14. S. Jitomirskaya and F. Yang, Singular Continuous Spectrum for Singular Potentials,
Commun. Math. Phys. {\bf 351}, 1127 (2017).\\
15. F. Yang and S. Zhang,  Singular Continuous Spectrum and Generic Full Spectral/Packing Dimension for Unbounded Quasiperiodic Schr\"odinger Operators,
Annales Henri Poincar\'e {\bf 20}, 2481 (2019).\\
16. T. Liu, X. Xia, S. Longhi, and L. Sanchez-Palencia, Anomalous mobility edges
in one-dimensional quasiperiodic models, SciPost Phys. {\bf 12}, 027 (2022).\\
17. Y.-C. Zhang and Y.-Y. Zhang, Lyapunov exponent, mobility edges, and critical region in the generalized
Aubry-Andre model with an unbounded quasiperiodic potential, Phys. Rev. B {\bf 105}, 174206 (2022).\\
18. S. Abe and H. Hiramoto, Fractal dynamics of electron wave packets in one-dimensional quasiperiodic systems, Phys. Rev. A {\bf 36}, 5349 (1987).\\
19. D.N. Christodoulides, F. Lederer, and Y. Silberberg, 
Discretizing light behaviour in linear and nonlinear waveguide lattices,
Nature {\bf 424}, 817 (2003).\\
20.  S. Longhi, Quantum-optical analogies using photonic structures,
Laser \& Photon. Rev. {\bf 3}, 243 (2009).\\
21. L. Yuan, Q. Lin, M. Xiao, and S. Fan,
Synthetic dimension in photonics, Optica {\bf 5}, 1396 (2018).\\
22.R. Morandotti, U. Peschel, J. S. Aitchison, H. S. Eisenberg, and Y. Silberberg, Experimental Observation of Linear and Nonlinear Optical Bloch Oscillations,
Phys. Rev. Lett. {\bf 83}, 4756 (1999).\\
23. N. Chiodo, G. Della Valle, R. Osellame, S. Longhi, G. Cerullo,
R. Ramponi, P. Laporta, and U. Morgner, Imaging of Bloch
oscillations in erbium-doped curved waveguide arrays, Opt.
Lett. {\bf 31}, 1651 (2006).\\
24. M. Wimmer, M.-A. Miri, D. Christodoulides, and U. Peschel,
Observation of Bloch oscillations
in complex PT-symmetric photonic
lattices, Sci. Rep.  {\bf 5}, 17760 (2015).\\
25. H. Chen, N. Yang, C. Qin, W. Li, B. Wang, T. Han, C. Zhang, W. Liu, K. Wang, H. Long, X. Zhang and P. Lu,
Real-time observation of frequency Bloch
oscillations with fibre loop modulation,  Light: Sci. \& Appl. {\bf 10}, 48 (2021).\\
26. A.V. Ponomarev and A.R. Kolovsky, Dipole and Bloch Oscillations of Cold Atoms
in a Parabolic Lattice, Laser Phys.  {\bf 16}, 367 (2006).\\
27. M.J. Zheng, Y.S. Chan, and K.W. Yu,
 Steering between Bloch oscillation and dipole
oscillation in parabolic optical waveguide arrays, J. Opt. Soc. Am. B. {\bf 27},  1299 (2010).\\
 28.M. Schreiber, Fractal eigenstates in disordered systems,
Physica A {\bf 167}, 188 (1990).\\
29. S. Longhi, Absence of mobility edges in mosaic Wannier-Stark lattices,
Phys. Rev. B {\bf 108}, 064206 (2023).\\
30. R. del Rio, S. Jitomirskaya, Y. Last, and B. Simon,
 What is Localization?,
Phys. Rev. Lett. {\bf 75}, 117 (1995).\\
 31. G.S. Jeon, B.J. Kim, S.W. Yi, and M.Y. Choi
 Quantum diffusion in the generalized Harper equation, J. Phys. A {\bf 31}, 1353. (1998).\\
 32. T. Xiao, D. Xie, Z. Dong, T. Chen, W. Yi, and B. Yan,
 Observation of topological phase with critical localization in a quasi-periodic lattice, Sci. Bull. {\bf 66}, 2175 (2021).\\ 
 33. S. Weidemann, M. Kremer, S. Longhi, and A. Szameit, Coexistence of dynamical delocalization and spectral localization through stochastic dissipation, Nature Photon. {\bf 15}, 576 (2021).\\
 34  H. Li, Z. Dong, S. Longhi, Q. Liang, D. Xie, and B. Yan,  Aharonov-Bohm Caging and Inverse Anderson Transition in Ultracold Atoms,
Phys. Rev. Lett. {\bf 129}, 220403 (2022).\\
35. A. Regensburger, C. Bersch, M. A. Miri, G. Onishchukov, D.N. Christodoulides, and U. Peschel,
Nature {\bf 488}, 167 (2012).\\
36. M. Wimmer, H.M. Price, I. Carusotto, and U. Peschel,
Experimental measurement of the Berry curvature from anomalous transport,
Nature Phys. {\bf 13}, 545 (2017).\\
37. A.L. M. Muniz, A. Alberucci, C.P. Jisha, M. Monika, S. Nolte, R. Morandotti, and U. Peschel,
Kapitza light guiding in photonic mesh lattice, Opt. Lett. {\bf 44}, 6013 (2019).\\
38. S. Wang, C. Qin, W. Liu, B. Wang,
F. Zhou, H. Ye, L. Zhao, J. Dong, X. Zhang,
S. Longhi, and  P. Lu,
High-order dynamic localization and tunable
temporal cloaking in ac-electric-field driven
synthetic lattices, Nature Commun. {\bf 13}, 7653. (2022).\\
39. S. Longhi, Non-Hermitian topological mobility edges and transport in photonic quantum walks, Opt. Lett. {\bf 47}, 2951 (2022).\\
40. M. Genske, W. Alt, A. Steffen, A.H. Werner, R.F. Werner, D. Meschede, and A. Alberti, Electric Quantum Walks with Individual Atoms,
Phys. Rev. Lett. {\bf 110}, 190601 (2013).\\
41. C. Cedzich, T. Ryb\'ar, A. H. Werner, A. Alberti, M. Genske, and R. F. Werner, Propagation of Quantum Walks in Electric Fields,
Phys. Rev. Lett. {\bf 111}, 160601 (2013).



\begin{thebibliography}{99}




\bibitem{r1}
P.W. Anderson,  Phys. Rev. {\bf 109},1492 (1958).
\bibitem{r2}
D.S. Wiersma, P. Bartolini, A. Lagendijk, and R. Righini, Nature {\bf 390}, 671 (1997).
\bibitem{r3}
T. Schwartz, G. Bartal, S. Fishman, and M. Segev,  Nature {\bf 446}, 52 (2007).
\bibitem{r4}
Y. Lahini, R. Pugatch, F. Pozzi, M. Sorel, R. Morandotti, N. Davidson, and Y. Silberberg, Phys. Rev. Lett.
{\bf 103}, 013901 (2009).
\bibitem{r5}
M. Segev, Y. Silberberg, and D.N. Christodoulides, Nature Photon. {\bf 7}, 197 (2013).
\bibitem{r6}
B. Simon and T. Spencer,  Commun.
Math. Phys. {\bf 125}, 113 (1989).
\bibitem{r7}
I. Kachkovskiy, J. Functional Analysis {\bf 277}, 3467 (2019).
\bibitem{r8}
D.B. Grempel, S. Fishman, and B. E. Prange, Phys. Rev. Lett. {\bf 49}, 833 (1982).
\bibitem{r9}
9. S. Fishman, D.B. Grempel, and B. E. Prange,  Phys. Rev.
Lett. {\bf 49}, 509 (1982).
\bibitem{r10}
B. Simon, Ann. Phys. {\bf 159}, 157 (1985).
\bibitem{r11}
B. Fischer, B. Vodonos, S. Atkins, and A. Bekker, Opt. Lett. {\bf 27}, 1061 (2002).
\bibitem{r12}
S. Ganeshan, K. Kechedzhi, and S. Das Sarma, Phys. Rev. B {\bf 90},
041405(R) (2014).
\bibitem{r13}
S. Longhi, Opt. Lett. {\bf 46}, 637 (2021).
\bibitem{r14}
S. Jitomirskaya and F. Yang,
Commun. Math. Phys. {\bf 351}, 1127 (2017).
\bibitem{r15}
F. Yang and S. Zhang,  Ann. Henri Poincar\'e {\bf 20}, 2481 (2019).
\bibitem{r16}
T. Liu, X. Xia, S. Longhi, and L. Sanchez-Palencia, SciPost Phys. {\bf 12}, 027 (2022).
\bibitem{r17}
Y.-C. Zhang and Y.-Y. Zhang, Phys. Rev. B {\bf 105}, 174206 (2022).
\bibitem{r17b}
S. Abe and H. Hiramoto,
Fractal dynamics of electron wave packets in one-dimensional quasiperiodic systems, Phys. Rev.  A {\bf 36}, 5349 (1987). 
\bibitem{r18}
D.N. Christodoulides, F. Lederer, and Y. Silberberg, Nature {\bf 424}, 817 (2003).
\bibitem{r19}
S. Longhi, Laser \& Photon. Rev. {\bf 3}, 243 (2009).
\bibitem{r20}
L. Yuan, Q. Lin, M. Xiao, and S. Fan, Optica {\bf 5}, 1396 (2018).
\bibitem{r21}
R. Morandotti, U. Peschel, J. S. Aitchison, H. S. Eisenberg, and Y. Silberberg, 
Phys. Rev. Lett. {\bf 83}, 4756 (1999).
\bibitem{r22}
 N. Chiodo, G. Della Valle, R. Osellame, S. Longhi, G. Cerullo,
R. Ramponi, P. Laporta, and U. Morgner,  Opt.
Lett. {\bf 31}, 1651 (2006).
\bibitem{r23}
M. Wimmer, M.-A. Miri, D. Christodoulides, and U. Peschel, Sci. Rep.  {\bf 5}, 17760 (2015).
\bibitem{r24}
H. Chen, N. Yang, C. Qin, W. Li, B. Wang, T. Han, C. Zhang, W. Liu, K. Wang, H. Long, X. Zhang and P. Lu,
 Light: Sci. \& Appl. {\bf 10}, 48 (2021).
 \bibitem{r25}
 A.V. Ponomarev and A.R. Kolovsky, Laser Phys.  {\bf 16}, 367 (2006).
 \bibitem{r26}
 M.J. Zheng, Y.S. Chan, and K.W. Yu, J. Opt. Soc. Am. B. {\bf 27},  1299 (2010).
 \bibitem{r27}
 M. Schreiber, Physica A {\bf 167}, 188 (1990).
 \bibitem{LonghiWS}
 S. Longhi, Phys. Rev. B {\bf 108}, 064206 (2023).
 \bibitem{r27b}
R. del Rio, S. Jitomirskaya, Y. Last, and B. Simon,
Phys. Rev. Lett. {\bf 75}, 117 (1995).
 \bibitem{r28}
 G.S. Jeon, B.J. Kim, S.W. Yi, and M.Y. Choi, J. Phys. A {\bf 31}, 1353. (1998).
 \bibitem{r29}
 T. Xiao, D. Xie, Z. Dong, T. Chen, W. Yi, and B. Yan, Sci. Bull. {\bf 66}, 2175 (2021). 
 \bibitem{r30}
 S. Weidemann, M. Kremer, S. Longhi, and A. Szameit, Nature Photon. {\bf 15}, 576 (2021).
 \bibitem{r30b}
 H. Li, Z. Dong, S. Longhi, Q. Liang, D. Xie, and B. Yan,
Phys. Rev. Lett. {\bf 129}, 220403 (2022).
 \bibitem{r31}
 A. Regensburger, C. Bersch, M. A. Miri, G. Onishchukov, D.N. Christodoulides, and U. Peschel,
Nature {\bf 488}, 167 (2012).
\bibitem{r32}
M. Wimmer, H.M. Price, I. Carusotto, and U. Peschel,
Nature Phys. {\bf 13}, 545 (2017).
\bibitem{r33}
A.L. M. Muniz, A. Alberucci, C.P. Jisha, M. Monika, S. Nolte, R. Morandotti, and U. Peschel,
 Opt. Lett. {\bf 44}, 6013 (2019).
\bibitem{r34}
S. Wang, C. Qin, W. Liu, B. Wang,
F. Zhou, H. Ye, L. Zhao, J. Dong, X. Zhang,
S. Longhi, and  P. Lu, Nature Commun. {\bf 13}, 7653. (2022).
\bibitem{r34b}
S. Longhi, Opt. Lett. {\bf 47}, 2951 (2022).
\bibitem{r35}
M. Genske, W. Alt, A. Steffen, A.H. Werner, R.F. Werner, D. Meschede, and A. Alberti, 
Phys. Rev. Lett. {\bf 110}, 190601 (2013).
\bibitem{r36} 
C. Cedzich, T. Ryb\'ar, A. H. Werner, A. Alberti, M. Genske, and R. F. Werner, Phys. Rev. Lett. {\bf 111}, 160601 (2013).





 








\end{thebibliography}
\end{document}